\documentclass[twocolumn,superscriptaddress,showpacs,nofootinbib,notitlepage,preprintnumbers,secnumarabic,amssymb, nobibnotes, prd]{revtex4-2}
\usepackage[utf8]{inputenc}
\usepackage[english]{babel}
\usepackage{graphicx}
\usepackage{blindtext}
\usepackage{amssymb}
\usepackage{amsmath}
\usepackage{graphicx}
\usepackage{tikz}
\usepackage{float}
\usepackage[braket, qm]{qcircuit}
\usepackage{hyperref}
\usepackage{xcolor}
\definecolor{winered}{rgb}{0.8,0,0}
\definecolor{darkb}{rgb}{0,0,0.8}
\hypersetup{
    colorlinks=true,
    citecolor=blue,
    linkcolor=winered,
    filecolor=red,      
    urlcolor=darkb,
}

\begin{document}
\preprint{ FERMILAB-PUB-22-002-SQMS-T}
\title{Noise Improvements in Quantum Simulations of sQED using Qutrits}
\author{Erik J. Gustafson}
\affiliation{Theory Division, Fermi National Accelerator Laboratory, Batavia, IL 60510, USA}
\date{\today}

\begin{abstract}
We present an argument for the advantages of using qudits over qubits for scalar Quantum Electrodynamics in $(1+1)$d. We measure the mass gap using an out of time correlator as a function of noise coming from an amplitude damping error channel and a generalized Pauli channel decoherence channel for both qubits and qutrits. For the same error in determination of the mass, the qutrit simulations can tolerate 10 to 100x larger gate noise than a qubit simulations. We find that 20 per-cent accuracy on the mass gap could be possible in the near future with a qutrits but is infeasible using qubits.  
\end{abstract}

\maketitle

\section{Introduction}
There are many successes  of Euclidean lattice QCD, LQCD, in the past decades for high energy physics. In particular these calculations have been able to provide determinations of static quantities such as hadron masses and decay rates \cite{davoudi2020, padmanath2019hadron}. However, LQCD has been unable to predict dynamical quantities such as inelastic scattering cross-sections above four particle inelastic thresholds and real-time dynamics due to sign problems\cite{Schafer:2009dj}. While recent work has been done to address these problems \cite{Alexandru:2016gsd, Meyer:2007dy,Aarts:2014nba,Hoshina:2020R+,2021PhRvD.104a4513K}, quantum computing provides an avenue for advantage over classical computing by circumventing this signal-to-noise problem entirely \cite{Feynman, Lloyd1073,Jordan:2011ci,Jordan_2018}. 

Real-time simulations of QCD in $3+1$ dimensions are decades away; quantum hardware is still too noisy and some necessary algorithmic tools are still missing.
Therefore, it is crucial to develop algorithms, regularization schemes, and resource requirements with simpler models that can inform future work.  \cite{Kogut:1979wt} proposed a road map that proved successful for LQCD.  This road map starts with simple $(1+1)d$ spin chains and gauge theories coupled to matter, both of which are already the focus of much study \cite{CerveraLierta2018,chertkov2021holographic,Scholl_2021,Labuhn_2016,PhysRevA.79.062314,yeteraydeniz2021scattering,vovrosh2020confinement,Kandala_2019,2017Natur.551..601Z,Kandala:2017aa,gustafson2019benchmarking,Lamm:2018siq,Salath__2015, 2017Natur.551..579B,Gustafson2021,Notarnicola_2020, Kokail:2018eiw, Klco:2018kyo,Martinez:2016yna,2021arXiv210713563Z} and proceeds through models of ever increasing complexity eventually culminating in LQCD in $(3+1)d$. 
The next part involves $(2+1)d$ Abelian gauges theories. Studies of the simplest model, $\mathbb{Z}_2$, are in progress as well \cite{gustafson2020quantum, Yamamoto_2020, homeier:2020dvw, multinode, Zohar:2016wmo, Davoudi:2019bhy, armon2021photonmediated}. $\mathbb{Z}_2$ gauge theories and Ising spin models are useful toy models to begin testing quantum computers, because their symmetries match nicely with widely available qubit hardware. However, for high energy physics, significant work toward best practices for more complicated Abelian and non-Abelian gauge simulations is another step that needs to be taken. 

While fermions can be easily mapped to a quantum computer \cite{JordanPWignerE, Bravyi_2002}, it is more difficult to map scalar fields and Abelian and non-Abelian gauge theories which possess continuous degrees of freedom. These theories need to be digitized and in order to map them to a quantum computer. Methods for digitizing for Abelian and non-Abelian gauge theories \cite{Unmuth_Yockey_2019, Bender_2020, Zohar_2012,Bazavov_2015,bazavov2015effective,PhysRevLett.121.223201,Unmuth_Yockey_2018, Gustafson_2021, Kaplan_2020, Raychowdhury_2020, kan2021lattice, Alexandru_2019, Hackett_2019,Ji_2020, ciavarella2021trailhead, klco_2020,Chandrasekharan:1996ih,Brower:1997ha,Beard_1998,BROWER2004149, kan2021lattice,2021arXiv211108015B} and scalar theories \cite{Jordan:2011ci, klco_2019,kurkcuoglu2021quantum,Barata_2021} are currently being actively studied. Abelian theories such as $U(1)$ can be approximated by the discrete group $\mathbb{Z}_N$. While any choice of $N$ can be used to approximate $U(1)$, in order to ensure the existence of the massless $U(1)$ phase at least $N\geq5$ is necessary if using naive actions \cite{CREUTZ1983149,PhysRevD.20.1915, PhysRevD.18.1174,PhysRevD.19.3698, PhysRevD.19.3715, PhysRevD.21.1013}, while improved actions let us use courser truncations \cite{Fukugita:1982kk}. Non-Abelian theories such as $SU(2)$ and $SU(3)$ can be approximated by crystal sub-groups \cite{Alexandru_2019, Hackett_2019,Ji_2020,PhysRevD.24.3319,Ji_2020,PhysRevD.22.2465,BHANOT1982337, PhysRevD.22.2465, PhysRevD.21.2316,Flyvbjerg:1984dj}. In many of the $U(1)$ approximations and all the discrete subgroup approximations the local Hilbert space for the link variables is not a power of 2. This becomes a problem for qubit based quantum computers; there are extra unused degrees of freedom on the qubits encoding the link that can become populated by quantum gate approximations or quantum noise. Both of these have the result of creating unphysical quantum states. In addition, most qubit based quantum computers today have a nearest neighbor connectivity that hampers studying models whose local Hilbert spaces have dimension larger than 2. This frequently requires the inclusion of swap gates. For this reason it is more desirable to have a quantum object or group of objects with a Hilbert space of the same dimension as the local Hilbert space of the gauge link. This can be addressed by qudits, a generalizations of qubits with $d$-states. 

Much work has been done on qubit based hardware since it is what is most widely available, however recent studies have shown that qudit (d-state quantum objects) based algorithms are frequently more efficient for quantum computing \cite{baker2020efficient,Gokhale_2019,2007PhRvA.75b2313R, 2015NatSR.514671G, Gustafson_2021,vargascalderon2021manyqudit,PhysRevA.101.022304, 2006circuitdecomp, 2020quditswap,PhysRevA.66.012303}. In addition there have been recent proof of principle implementations of qutrits \cite{2010PhRvL.105v3601B,morvan2020qutrit, blok:2020may,2021arXiv210405627C, ringbauer2021universal,vashukevich2021highfidelity, 2020practicalquditTI,2018photonicqudits} and cavity QED for qubit-qudit couplings \cite{PhysRevLett.100.060504,2015snapgate, 2015qutrittransfer, 2010NatPh6772N,forndiaz2010observation,2004cqedChiorescu, 2004cqedWallraff}. In the context that this paper examines the benefits of qubit versus qutrit encodings for noisy quantum computers using $(1+1)d$ scalar QED, sQED, as a first step. 

A gauge invariant representation of this model was initially studied in \cite{bazavov2015effective,Bazavov_2015,PhysRevLett.121.223201,Unmuth_Yockey_2018}. This work extends the study done in \cite{Gustafson_2021} by looking more closely at emulations of NISQ era machines to simulate sQED. In order to study this comparison we will measure the mass gap as an example to show how qutrits are better for quantum simulations than qubits.

This paper is organized as follows. Sec. \ref{sec:Theory} lays out background theory and Hamiltonian for sQED and the observable that will be measured.  Sec. \ref{sec:Costs} outlines the circuit costs and the set of native gates used in this paper. Sec. \ref{sec:Errors} briefly outlines the two quantum noise models, and the results of using these noise models with strengths across multiple orders of magnitude for qubit and qutrit machines are discussed in Sec. \ref{sec:Results}.
Sec. \ref{sec:Conclusions} closes out this work and discusses the outlook for qubit versus qutrit based quantum computers for simulating sQED on a digital quantum computer.
\section{Theory}
\label{sec:Theory}
Following references \cite{bazavov2015effective,Bazavov_2015,PhysRevLett.121.223201,Unmuth_Yockey_2018} and summarizing \cite{Gustafson_2021}, the Euclidean action for the sQED in $(1+1)$d is
\begin{equation}
\label{eq:abelianhiggs}
\begin{split}
\mathcal{S} = & -\frac{1}{g^2 a_s a_{\tau}} \sum_x \sum_{\nu < \mu}\text{ReTr}(U_{x,\mu\nu})\\
&-\frac{a_\tau}{a_s} \sum_{x}\Big( \phi^{\dagger}_x U_{x,s} \phi_{x+\hat{s}} + h.c.\Big)\\
    & -\frac{a_s}{a_\tau} \sum_{x}\Big( \phi^{\dagger}_x U_{x,\tau} \phi_{x+\hat{\tau}} + h.c.\Big).
\end{split}
\end{equation}
The bare mass of the fundamental scalars is an additive constant because the magnitude of the scalars has been fixed to unity. In Eqs. (\ref{eq:abelianhiggs}), $g^2$ is the gauge coupling, $a_s$ is the spacial lattice spacing, and $a_t$ is the temporal lattice spacing. The compact fields $U_{x,\mu}$ and $\phi_{x}$ are defined as
\begin{equation}
\label{eq:fields}
U_{x, \mu} = e^{i A_{x, \mu}} \text{ and } \phi_{x} = e^{i \theta_x}.
\end{equation}

The derivation of the Hamiltonian is carried out in detail  \cite{bazavov2015effective,Bazavov_2015,PhysRevLett.121.223201,Unmuth_Yockey_2018}. The derivation involves integrating out the matter fields. Using tensorial methods the transfer matrix is used to take the time continuum limit. In this limit we end up with the rotor Hamiltonian,
\begin{equation}
\label{eq:hamiltonian}
\hat{H} = \hat{H}_V + \hat{H}_K
\end{equation}
where
\begin{equation}
\label{eq:h1}
    \hat{H}_V = \frac{g^2 a_s}{2} \sum_{i = 1}^{N_s} (\hat{L}^z_i)^2 + \frac{1}{2a_s} \sum_{i = 1}^{N_s}(\hat{L}_i^z - \hat{L}_{i + 1}^z)^2
\end{equation} is the gauge and potential terms,
\begin{equation}
\label{eq:h2}
     \hat{H}_K = - \frac{2}{a_s} \sum_{i = 1}^{N_s} \hat{U}^x_i,
\end{equation}
are the matter dynamics terms,
and the effect of the operators $\hat{L}^z$ and $\hat{U}^x$ on the rotor states $|n\rangle$ are
\begin{equation}
\label{eq:operators1}
\hat{L}^z_i |n\rangle_i =  n |n\rangle_i
\end{equation}
and 
\begin{equation}
    \label{eq:operators2}
\hat{U}^x_i|n\rangle_i = \frac{1}{2}\big(|n - 1\rangle_i + |n+1\rangle_i\big).
\end{equation}
The Hilbert space operators in Eqs. (\ref{eq:operators1}) and (\ref{eq:operators2}) act on are formally infinite dimensional.
In this work we will work in lattice units where $a_s = 1$. In addition we choose $g^2 a_s=5$ so that the effects of higher order states are negligible due to the $(\hat{L}^z)^2$ term making higher rotor states less energetically favorable  \cite{Gustafson_2021, Unmuth_Yockey_2018}.

Up to this point the Hilbert space the operators in Eq. (\ref{eq:hamiltonian}) act on is infinite. In practice this Hilbert space needs to be truncated so that it is practical to implement on a quantum computer due to memory constraints. The structure of the operators in this model are naturally truncated to an odd-dimensional local Hilbert space. This is to avoids an asymmetry in quantum states of the links, \textit{i.e.} the states range from $-m$ to $m$ rather than $-m + 1$ to $m$, where $m$ is the maximal value allowed on a link. For this reason the simplest mapping for this model would require 3 states. If we consider trunctating the $U(1)$ symmetry to three states, we can map this Hamiltonian onto two qubits or one qutrit. The allowable quantum states on the given link are $|\pm1\rangle$ and $|0\rangle$. While the work that follows uses qutrits, the same methodology can be applied to higher dimensional operators as well. 
In this way, the operators in Eq. (\ref{eq:operators1}) and Eq. (\ref{eq:operators2}) can be written as the following matrices:
\begin{equation}
\label{eq:lzqutrit}
\hat{L}^z_i = \begin{pmatrix} 1 & 0 & 0\\ 0 & 0 & 0\\ 0 & 0 & -1\\ \end{pmatrix}
\end{equation}
and 
\begin{equation}
\label{eq:uxqutrit}
\hat{U}^x_i = \frac{1}{2} \begin{pmatrix} 0 & 1 & 0 \\ 1 & 0 & 1\\ 0 & 1 & 0\end{pmatrix}.
\end{equation}
For a quantum simulation of a lattice field theory we need to have some way to determine what the actual lattice spacing is. If we want to extract continuum physics, it is important to know what the lattice spacing is. Methods for this in regard to quantum simulations are currently being studied \cite{carena2021lattice}. In order to determine the lattice spacing typically the mass of some particle of interest is used. This mass in Euclidean calculations is typically extracted using a temporal correlator,
\begin{equation}
\label{eq:correlator}
\mathcal{C}(t) = \langle O(t) O(0)\rangle = \langle \psi| \hat{\mathcal{U}}^{\dagger}(t) \hat{O} \hat{\mathcal{U}}(t) \hat{O}|\psi\rangle.
\end{equation}

While \cite{carena2021lattice} indicated it may not be necessary to use quantum simulations to set the scale, calculations of many different dynamical quantities will be effected by the noise of quantum computers and using a real-time correlator will provide a pessimistic estimate of the resources required to extract results from a quantum calculation. 

To measure the extract the mass of the lightest bound state we need to measure the correlator,
\begin{equation}
\label{eq:csQEDcor}
\mathcal{C}(t) = \langle \Gamma| e^{i t \hat{H}} \sum_{x} \hat{U}^{-}_x e^{-i t \hat{H}} U^+_0 |\Gamma\rangle.
\end{equation}
The operators $\hat{U}^\pm$ raise or lower the state $|n\rangle$ by one.  $|\Gamma\rangle$ is a state that has a roughly good overlap with the ground state when taken as a tensor product over all the sites,
\begin{equation}
\label{eq:gammastate}
|\Gamma\rangle = (|-1\rangle + b |0\rangle + |1\rangle)/\sqrt{\mathcal{N}},
\end{equation} 
where
\begin{equation}
    b = \frac{g^2 + 1 + \sqrt{(g^2 - 1)^2 + 32}}{4}
\end{equation}
and
\begin{equation}
    \mathcal{N} = \sqrt{2 + b^2}.
\end{equation}
This state is chosen for the following reason. Since the interaction term is is weak the ground state can roughly be approximated by diagonalizing the operator
\begin{equation*}
    \Big(\frac{g^2 + 1}{2} +1 \Big)(\hat{L}^z)^2 - 2 \hat{U}^x.
\end{equation*} 
Since $|\Gamma\rangle$ roughly approximates the ground state,
\begin{equation}
    \label{eq:gammaevolve}
    e^{-i t \hat{H}} |\Gamma \rangle \sim e^{-i t E_0} |\Gamma\rangle,
\end{equation}
where $E_0$ is the smallest eigenvalue of Eq. (\ref{eq:hamiltonian})
Similarly since $\hat{U}^+$ will roughly excite the lightest bound state \cite{Gustafson_2021}, 
\begin{equation}
    \sum_{x} U^-_x e^{-i t \hat{H}} \hat{U}^+_0|\Gamma\rangle \sim e^{-i t E_1} |\Gamma\rangle.
\end{equation}
$E_1$ is the eigenvalue of the lightest bound state. Combining these two equations together will have Eq. (\ref{eq:correlator}) given by
\begin{equation}
\mathcal{C}(t) \sim e^{i t (E_0 - E_1)} = e^{-i t m},
\end{equation} 
where $m$ is the mass of the particle. We will extract this energy difference by taking a fast Fourier transform of the time series correlator to generate a frequency spectrum. The mass will correspond to the largest peak in the frequency spectrum.

In the following section we discuss the systematic errors that will arise from using the correlator to set the scale $\delta t$ for a quantum simulation.

\section{Systematics}
\label{sec:Systematics}
There are many systematic errors that are noticeable for any NISQ era QFT simulation. These systematic errors will come from two places: the Fourier transform of the time series data, and Trotterization. These first two types of errors are also going to exist in fault tolerant computers, while systematic and coherent errors from gate operations will only be a NISQ era problem.

Given the corrrelator in Eq. (\ref{eq:csQEDcor}), a Fourier transform will take this time series data and provide an energy spectrum that should be sharply peaked around the desired state of interest. In practice this is done by taking a finite number of time steps at a finite resolution of $\delta t$. The lower bound on the uncertainty is therefore going to be
\begin{equation}
\label{eq:dE_FFT}
\delta E_{FFT} = 2 \pi / (\delta t N_{steps}),
\end{equation}
where $\delta t$ is time resolution and $N_{steps}$ is the number of sample in time. The time resolution, $\delta t$, directly imposes an upper bound on the observable energy. $N_{steps}$ imposes a finite resolution on the energy given a choice of $\delta t$. 
It should be reiterated, the resolution in Eq. (\ref{eq:dE_FFT}) is a lower bound. In practice the imprecision of the source operators $\hat{U}^\pm$ will impose ancillary excitations and decreased resolution of the peaks. 

In practice implementing $e^{-i t \hat{H}}$ is impractical and we will need to resort to Trotterization to implement the time evolution. This involves approximating $e^{-i t \hat{H}}$ by

\begin{equation}
    \label{eq:trotterization}
    e^{-i\delta t\hat{H}} \sim \hat{\mathcal{U}}_{Tr}(\delta t) = \hat{\mathcal{U}}_z(\delta t)\hat{\mathcal{U}}_{zz}(\delta t)\hat{\mathcal{U}}_x(\delta t)
\end{equation}
where
\begin{equation}
    \label{eq:Uz}
   \hat{\mathcal{U}}_z(\delta t) = e^{-i \delta t (g^2 / 2 + 1)\sum_i(\hat{L}^z_i)^2},
\end{equation}
\vskip-2em
\begin{equation}
\label{eq:Uzz}
    \hat{\mathcal{U}}_{zz}(\delta t) = e^{-i \delta t\sum_i(\hat{L}^z_i\hat{L}^z_{i + 1})},
\end{equation}
and
\begin{equation}
\label{eq:Uxevo}
    \hat{\mathcal{U}}_x(\delta t) = e^{-i \delta t\sum_i(\hat{U}^x_i)}.
\end{equation}
The Trotterization of the time evolution operator is going to take us away from the correlator in Eq. (\ref{eq:csQEDcor}) to
\begin{equation}
\label{eq:trotter}
	\mathcal{C}(t; n) = \langle \Gamma| e^{i t \hat{H}_{BCH}} \sum_{x} U^{-}_x e^{-i t \hat{H}_{BCH}} U^-_0|\Gamma\rangle.
\end{equation}
The term, $\hat{H}_{BCH}$, is the Baker-Campbell-Hausdorff (BCH) Hamiltonian that is actually being simulated via Trotterization \cite{carena2021lattice}. This Hamiltonian is
\begin{equation}
\label{eq:BCHham}
\begin{split}
\hat{H}_{BCH} = & \hat{H} - \frac{\delta t^2}{24} \Big([2\hat{H}_K [\hat{H_K}, \hat{H}_V]]\\
& + [\hat{H}_V,[\hat{H}_V,\hat{H}_K]]\Big) + \mathcal{O}(\delta t^3),
\end{split}
\end{equation}
where $\hat{H}_1$ and $\hat{H}_2$ are defined in Eq. (\ref{eq:h1}) and Eq. (\ref{eq:h2}) respectively.
As a consequence, Trotterization will impose an $\mathcal{O}(\delta t^2)$ systematic error on the energies measured from the correlator. 

\begin{figure}
\includegraphics[width=0.45\textwidth]{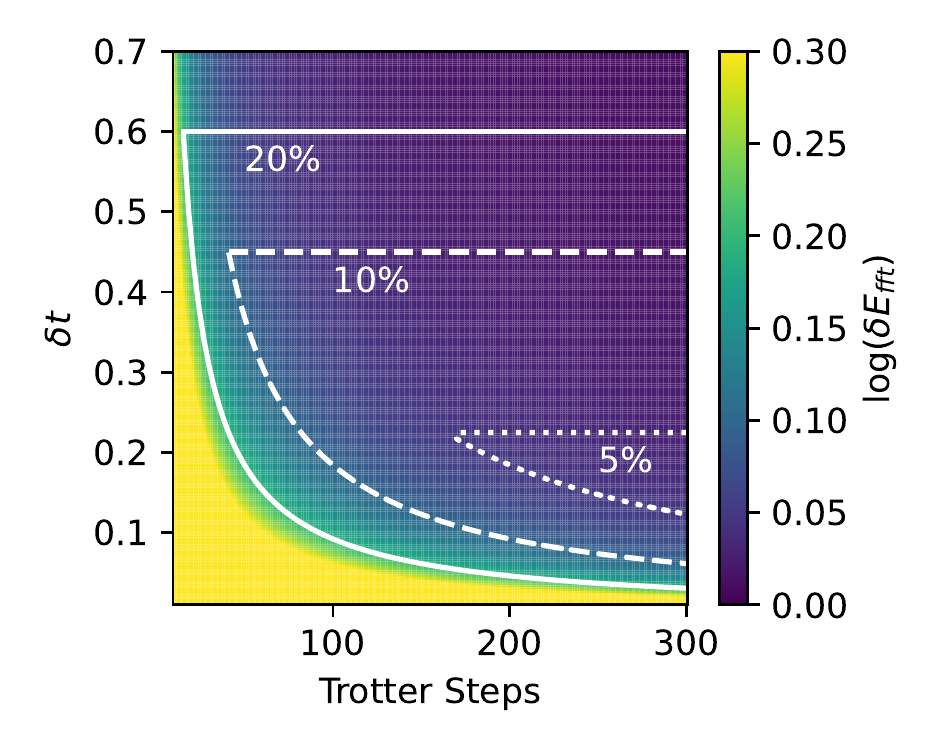}
\caption{Systematic error from Fourier transform as a function of the number of Trotter steps and Trotter step size. The solid, dashed, and dotted white lines bound the region where the accuracy and precision are sub 20, 10  and 5 per-cent accuracy and precision.}
\label{fig:accuracybounds}
\end{figure}

The uncertainty from the Fourier transform is plotted in Fig. \ref{fig:accuracybounds}. In addition the regions corresponding to certain chosen accuracies and precisions (20, 10, and 5 per-cent) that are bounded both by the Fourier transform and the BCH Hamiltonian. 
These bounds are summarized in Tab. \ref{tab:sysreq}. The minimal number of Trotter steps increases quickly with respect to the desired accuracy. Similarly the required Trotter step size, $\delta t$, shrinks quickly as well. 
\begin{table}[ht]
    \centering
    \begin{tabular}{ccc}
    \hline\hline
         Accuracy & $\delta t < $ & Min. $N_{steps}$  \\\hline
         $20\%$ & 0.6 & 20 \\
         $10\%$ & 0.36 & 50 \\
         $5\%$ & 0.25 & 180 \\\hline\hline
    \end{tabular}
    \caption{Maximal $\delta t$ and minimum $N_{steps}$ to achieve a given systematic accuracy on the mass gap. These bounds are derived from Fig. \ref{fig:accuracybounds}}
    \label{tab:sysreq}
\end{table}

\section{Circuits}
\label{sec:Costs}

Qutrit circuits do not have a unique mapping onto qubit circuits. Assuming that linear combinations for the encoding are not used, there are twelve different different ways one can encode a qutrit state onto a pair of qubits. In this work we use the mapping,
\begin{equation}
\begin{split}
|0\rangle_3 = & |00\rangle_2, ~|1\rangle_3 = |01\rangle_2, ~|2\rangle_3 =  |10\rangle_2,\\
\end{split}
\end{equation}
although the end choice should not matter dramatically. 
The subscripts here denote whether the state is a qubit, $|\rangle_2$, or a qutrit $|\rangle_3$ state. A side effect of this mapping is that the rotations between the states $|0\rangle_3$, $|1\rangle_3$, and $|2\rangle_3$ do not need an entangling gate but the qubit states need entangling operations to rotate between the states, $|01\rangle_2$ and $|10\rangle_2$. In addition encoding a qutrit onto a qubit has an added difficulty that an unused fourth state $|11\rangle_2$ remains. Including this state results in the un-physical portions of the Hilbert space being reached when gate noise is present that will allow coupling between the $|11\rangle$ state and the other states. This is nearly identical to the problem in quantum simulation of gauge theories, where noise can move the simulation out of the physical Hilbert space \cite{2020arXiv200512688L,2021arXiv211008041V,2021arXiv210802203H,2020arXiv200907848H,2021PRXQ....2d0311H}. There are ways to minimize these errors on noisy quantum computer using proceedures such as Pauli Twirling and randomized compiling \cite{Erhard_2019, 2018efficienttwirling, 2013efficienttwirl, 2016efficienttwirling}. However Pauli Twirling and randomized compiling may still not address the issue of populating the unphysical state entirely. Because Pauli twirling transforms an error channel into a stochastic one, random errors populating the unphysical state could still be present or amplified because there will always be a non-zero probability of producing the $|11>$ state. 

Two important, but not exclusive, factors when considering these two encodings for NISQ devices are: entangling gate depth and idle time. Entangling gates are typically the longest and noisiest operation on a NISQ device, therefore minimizing entangling gate depth is crucial for any quantum simulation.  Idle time is also a major contributing factor. When qudits are idling the excited state populations can dephase and decay. This is a greater issue with qudits because they higher energy states decay faster by 25 to 50 $\%$ \cite{morvan2020qutrit,blok:2020may}. This decay can be mitigated in some ways by dynamic decoupling \cite{PhysRevLett.82.2417,1999PhLA..258...77Z,2021QuantumVolume}. It is worth noting that cross-talk \cite{Gokhale_2019, Erhard_2019,Sarovar_2020, 2016PhRvA..93f0302S, PRXQuantum.1.020318,2021npjQI...7..129H}  is an additional noteworthy source of noise in many quantum computers but in transmon  systems due to frequency crowding \cite{PRXQuantum.1.020318,2021npjQI...7..129H} but is a topic for later investigation as it involves non-local and spectator errors that make the noise models significantly more complicated.

The quantum circuits required can be split into two parts: state preparation and Trotterization. For qubits, the primitive gates are assumed to be parameterized $R_x$, $R_y$, and $R_z$ rotations and the controlled not gate. For qutrits, the primitive gates are $R_x$, $R_y$, and $R_z$ rotations in the the $|0\rangle$- $|1\rangle$ and $|1 \rangle$-$|2\rangle$ subspace, and the controlled sum gate, 
\begin{equation}
    \label{eq:csumgate}
    CSUM = \sum_{i = 0}^{2}\sum_{j=0}^{2} |i\rangle\langle i|\otimes |j\rangle \langle (j + i) mod~3|.
\end{equation}
The CSUM gate is a natural extension of the CNOT to a qutrit. A summary of all the gate costs is provided in Tab. \ref{tab:gatecostssummary}. 

\begin{table}
\begin{tabular}{ccc|cc}
\hline
\hline
Operator & 1-qubit & 2-qubit & 1 qutrit & 2-qutrit\\
\hline
$V_g$    & 3 & 2 & 2 & 0\\
$CU_g$  & 54 & 54 & 5 & 2\\
$e^{-i\theta (L^z)^2}$ & 1 & 0 & 2 & 0 \\
$e^{-i\theta U^x}$ & 6 & 2 & 5 & 0 \\
$e^{-i\theta L^zL^z}$ & 4 & 26 & 4 & 3 \\
\hline
\hline
\end{tabular}
\caption{Basic gate costs for the qubit and qutrit encodings of each operator in the quantum circuit.}
\label{tab:gatecostssummary}
\end{table}

\begin{figure}[!ht]
\centering
\includegraphics[width=3.375in]{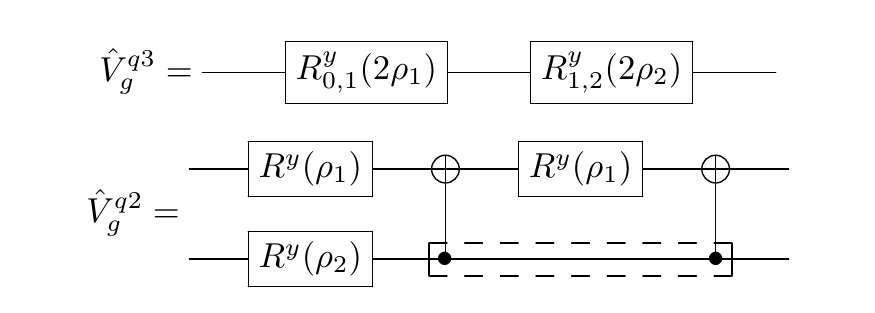}
\caption{Quantum Circuit to carry out the $V_g$ operator on a qutrit (top) and qubit register (bottom) to create the one qutrit state $|\Gamma\rangle$ from Eq. (\ref{eq:gammastate}). The angles $\rho_1$ and $\rho_2$ are provided in Eqs. (\ref{eq:qutritvgangles}) and (\ref{eq:qubitvgangles}). The dashed box indicates a region where the qubits are idling.}
\label{fig:stateprepcircuits}
\includegraphics[width=3.375in]{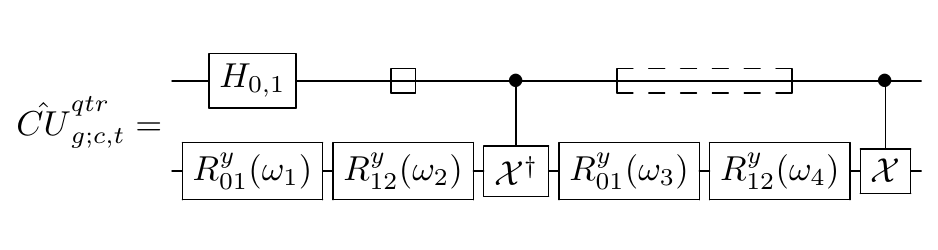}
\caption{The quantum circuit to implement the operation to create the superposition state for the source operator $\hat{CU}_g$. Which implements creating the initial state $\hat{U}^+|\Gamma\rangle + |\Gamma\rangle$. These create the states used to measure correlator in Eq. (\ref{eq:csQEDcor}). The qubit circuit is long and determined using QISKit. The boxed areas of the circuit indicate times when the qutrit is idling. The $c$ and $t$ subscripts indicate the control (top line) and target (bottom line) qutrits.}
\label{fig:stateprepcircuits2}
\end{figure}

The state preparation circuits take the quantum system from the initial computational qutrit state to the desired initial state. There are two operators that we need for this state preparation,
$\hat{V}_g$ which takes $|0\rangle$ to $|\Gamma\rangle$ and $\hat{CU}$ which takes the $|0\rangle$ state to $|\Gamma\rangle|0\rangle_a + \hat{U}^+|\Gamma\rangle|1\rangle_a$, where the $_a$ indicates the use of an ancilla qubit / qutrit which is needed to measure a unitary observable. 
In Fig. \ref{fig:stateprepcircuits} are the qutrit implementations of the $\hat{V}_g$ operator from \cite{Gustafson_2021}. The angles $\rho_1$ and $\rho_2$ are given by the following equations,
\begin{equation}
\label{eq:qutritvgangles}
\begin{split}
\rho_1 =& -\arccos(1 / \mathcal{M})\\
\rho_2 =& -\arccos(1 / \sqrt{\mathcal{M}^2 - 1})\\
\mathcal{M} =& \sqrt{2 + \frac{1 + 2 g^2 + \sqrt{129 + 4 g^2 + 4 g^4}}{8}}.
\end{split}
\end{equation} 
The first rotation create a superposition of $|0\rangle$ and $|1\rangle$ state. The second angle takes the $|1\rangle$ state and rotates it so that there is an equal probability in the $|0\rangle$ and $|2\rangle$ states. 
The qutrit encoding of this operator has two NISQ-era advantages over the qubit encoding. The first advantage is the qutrit encoding requires no entangling gates while the qubit encoding requires two controlled not gates. Secondly, qutrits in contrast to qubits require no idle time to perform this circuit.

\begin{table}[!ht]
\begin{tabular}{ccccc}
\hline\hline
angle & $\omega_1$ & $\omega_2$ & $\omega_3$ & $\omega_4$\\\hline
value & 1.5902 & 1.9847 & 2.4373 & 1.5911\\\hline
\end{tabular}
\caption{Optimal angles for the $R^y_{i,j}$ qutrit rotations shown in Fig. \ref{fig:stateprepcircuits2} for the given choice of $g$.}
\label{tab:omegasqutrit}
\end{table}
We perform the excitation operator by applying a controlled unitary $\hat{CU}_g^{qtr}$ which carries out the following transformation
\begin{equation}
\label{eq:cugtr}
    \hat{CU}_g^{qtr}|0\rangle(|0\rangle_a + |1\rangle_a) = |\Gamma\rangle|0\rangle_a + \frac{1}{\mathcal{N}'}U^+|\Gamma\rangle|1\rangle_a.
\end{equation} The subscript $a$ indicates it is an ancilla qutrit state. The ancilla qutrit state is used in conjunction with $\hat{U}^-$ at the end of the circuit to measure the correlator in Eq. (\ref{eq:csQEDcor}). The quantum circuit to encode the source excitation in the qutrit encoding is provided in Fig. \ref{fig:stateprepcircuits2}. The angles $\omega_i$ are Hamiltonian coefficient dependent and found by the unitary of the form in Fig. \ref{fig:stateprepcircuits2} that creates the state in Eq. (\ref{eq:cugtr}). These angles are listed in Tab. \ref{tab:omegasqutrit}.

The key features of this encoding are that only two two-qutrit entangling gates are necessary and there is a single 1-qutrit rotation of idle time on the ancilla qutrit. The qubit encoding is sixteen times longer in terms of entangling gate depth and is provided in Fig. \ref{fig:qubitcug} in the appendix. It was  derived using the unitary decomposition tool in QISKit \cite{Qiskit}. Provided an all to all connectivity this gate would require 15 CNOTs which is slightly above the theoretical lower bound for an arbitrary 3-qubit gate (13 CNOTs) and below many optimized compiling methods (21 CNOTs) \cite{2021arXiv210102993K, 2004PhRvL..93m0502M}. While this encoding seems close to optimal, if the qubits are connnected along a line then this CNOT cost rises to 51 due to the inclusion of swap gates. This is troublesome because 36 of these CNOTs are part of swap operations. In addition because of the number of the entangling gate operations there are 48 CNOT gates where the other 6 qubits are idling. 

\begin{figure}[!th]
\includegraphics[width=3.375in]{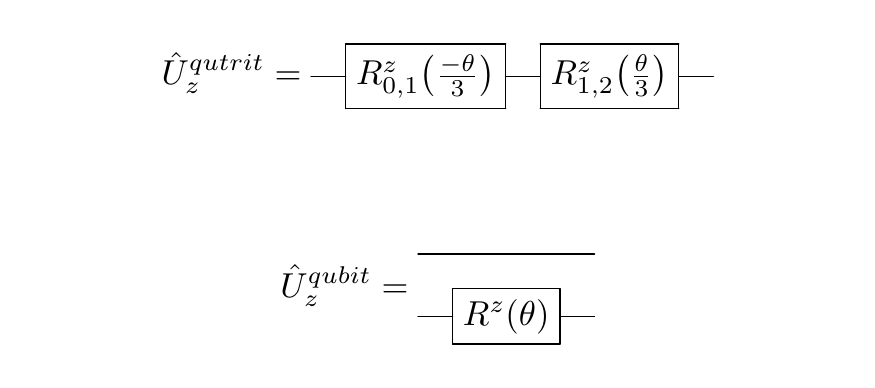}
\caption{Quantum Circuits to carry out the $e^{-i \theta (\hat{L}^z)^2}$ operator on qutrits (top) and qubits (bottom) in Eq. \ref{eq:Uz}.}
\label{fig:lz2rotationcircuit}
\end{figure}

The second set of circuits involve the Trotterization of the Hamiltonian. These can be separated into the three rotations given in Eqs. (\ref{eq:Uz}), (\ref{eq:Uzz}), and (\ref{eq:Uxevo}).
The qubit and qutrit circuits for Eq. (\ref{eq:Uz}) do not require an entangling gate and are in general noise free due to the hardware implementations \cite{Qiskit,blok:2020may,morvan2020qutrit}. Graphical depictions of these circuits are provided  are provided in Fig. \ref{fig:lz2rotationcircuit}.

\begin{figure*}
\includegraphics[width=6.75in]{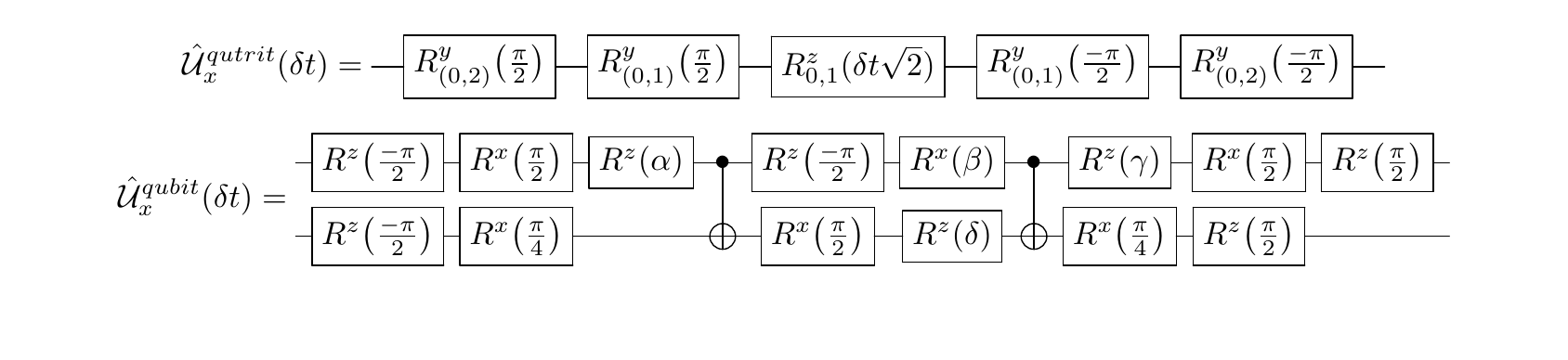}
\caption{Quantum circuit to implement $e^{i \theta U^x}$ rotation from Eq. (\ref{eq:trotterization}) on qutrits (top) and qubits (bottom). $\alpha$, $\beta$, $\gamma$, $\delta$ are all angles that are solely functions of $\delta t$.}
\label{fig:uxrotationcircuits}

\includegraphics[width=6.5in]{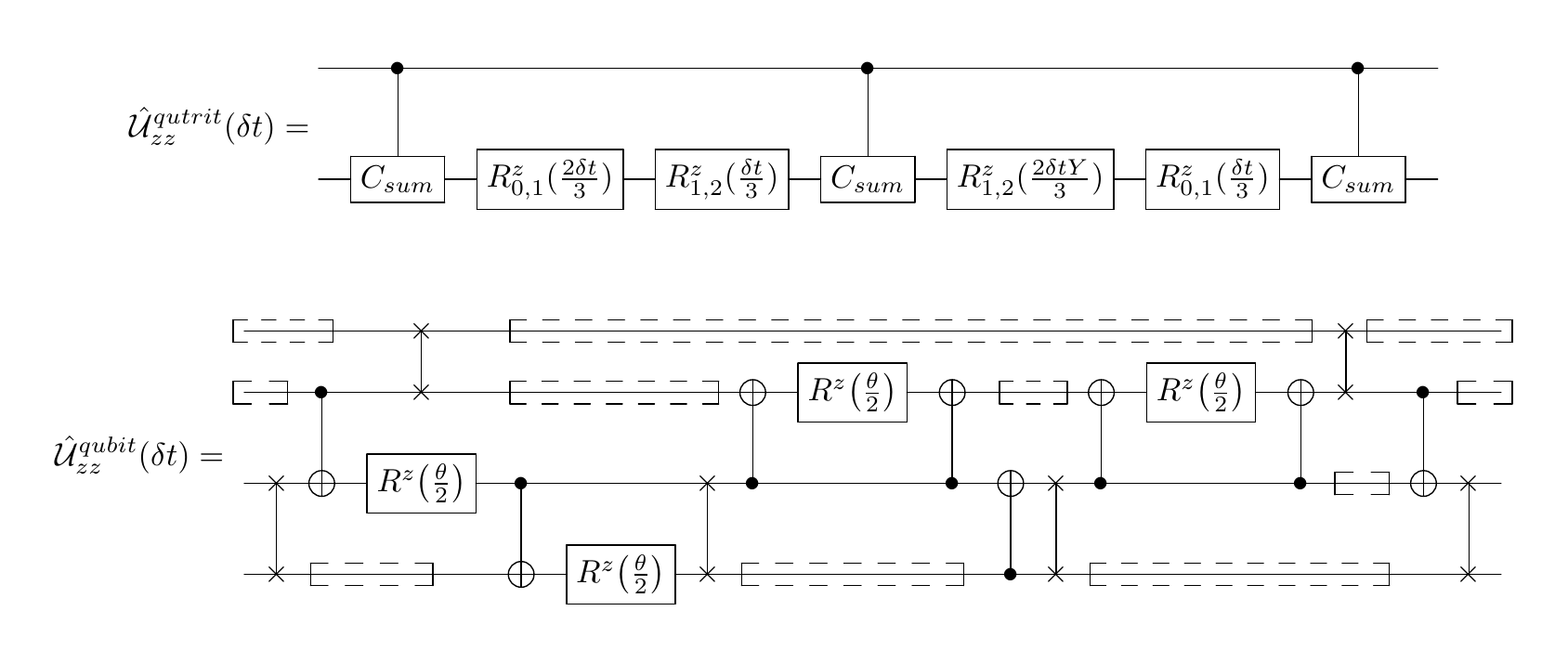}
\vskip1em
\begin{equation*}
\hat{\mathcal{U}}^{qubit}_{zz}(\delta t) = 
\begin{gathered}
\Qcircuit @R=1em @C=1em {
& \qw & \qw & \qswap & \qw & \qw & \qw & \qw & \qw & \qw & \qw & \qw & \qw & \qw & \qw & \qswap & \qw & \qw & \qw & \qw  \\
& \qw & \ctrl{1} & \qswap \qwx& \qw & \qw & \qw & \targ & \gate{R^z\big(\frac{\theta}{2}\big)} & \targ & \qw & \qw & \targ & \gate{R^z\big(\frac{\theta}{2}\big)} & \targ & \qswap \qwx & \qw & \ctrl{1} & \qw & \qw  \\
& \qswap & \targ & \gate{R^z\big(\frac{\theta}{2}\big)} & \ctrl{1} & \qw & \qswap & \ctrl{-1} & \qw & \ctrl{-1} & \targ & \qswap & \ctrl{-1} & \qw & \ctrl{-1} & \qw & \qw & \targ & \qswap & \qw \\
& \qswap \qwx & \qw & \qw & \targ & \gate{R^z\big(\frac{\theta}{ 2}\big)} & \qswap \qwx & \qw & \qw & \qw & \ctrl{-1} & \qswap \qwx & \qw & \qw & \qw & \qw & \qw & \qw & \qswap \qwx & \qw 
\gategroup{1}{1}{1}{3}{0.7em}{--}
\gategroup{4}{3}{4}{4}{0.7em}{--}
\gategroup{2}{1}{2}{2}{0.7em}{--}
\gategroup{2}{5}{2}{7}{0.7em}{--}
\gategroup{4}{8}{4}{10}{0.7em}{--}
\gategroup{2}{11}{2}{12}{0.7em}{--}
\gategroup{4}{13}{4}{17}{0.7em}{--}
\gategroup{3}{16}{3}{17}{0.7em}{--}
\gategroup{1}{5}{1}{15}{0.7em}{--}
\gategroup{1}{17}{1}{20}{0.7em}{--}
\gategroup{2}{19}{2}{20}{0.7em}{--}
}
\end{gathered}
\end{equation*}
\caption{Quantum circuits for implementing the $\hat{L}^z\otimes\hat{L}^z$ from Eq. (\ref{eq:trotterization}) rotations on qutrits (top) and qubits (bottom). For the qubits the two lines correspond to the top line on the qutrit circuit and the bottom two lines correspond to the bottom line on the qutrit circuit. The boxes indicate all the places where qubits are idling (no gate operations are being applied).}
\label{fig:lzlzrotationcircuit}
\end{figure*}
\begin{figure}[ht]
\includegraphics[width=3.375in]{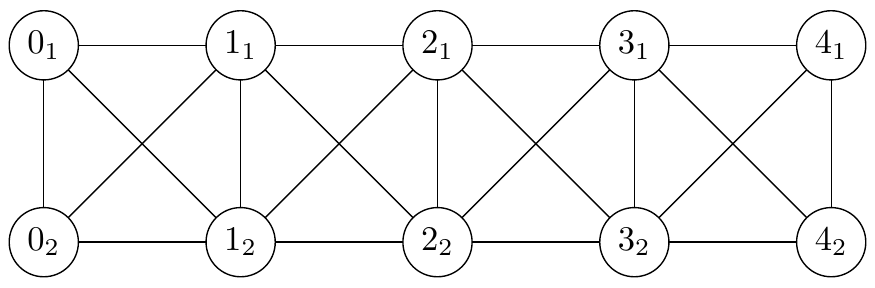}
%
%
\caption{Qubit layout to eliminate need for swap gates in qutrit encoding of $(1+1)d$ sQED. The circles denote qubits, the numbers within denote the different rotors and the subscript indicates the qubit composing the qutrits. The lines connecting circles indicate the which qubits are connected.}
\label{fig:qubitlayout}
\end{figure}
Fig. \ref{fig:uxrotationcircuits} shows the rotations for implementing the $e^{-i \delta t U^x}$ gate in both encodings.
The angles for the the qubit encoding are less straight forward to determine and are found by using the QISKit unitary decomposition. These angle are given by
\begin{equation}
\label{eq:qubitvgangles}
\begin{split}
\alpha =& f(\delta t)\\
\beta =& \pi - \pi / 2 \sin(\delta t /(2\sqrt{2}))\\
\gamma = & \pi / 2 - \alpha\\
\delta = & - \frac{\delta t}{2\sqrt{2}}\\
\end{split}
\end{equation}
where $f(\delta t)$ does not appear to have a simple closed form solution but appears to be composed of a linear term with oscillatory corrrections.
In both encodings there is no idle time. The substantive difference is that entangling gates are not necessary for the qutrit encoding.

Fig. \ref{fig:lzlzrotationcircuit} shows the implementation of the coupling term $e^{-i \theta \hat{L}^z\hat{L}^z}$. Since the $R^z$ rotations are implemented by reference frame changes on the quantum hardware there is no idle time on the qutrits. However in addition there 8 required entangling gates and 6 swap (18 controlled not) gates for the qubit encoding while the qutrit encoding requires 3 controlled sum gates. The qutrit encoding also is nearest neighbor while the qubit encoding requires swaps. If instead the qubits that compose the synthetic qutrit are connected to all the qubits that compose the neighboring synthetic qutrits then the swaps can effectively be eliminated an illustration of this connectivity is provided in Fig. \ref{fig:qubitlayout}.

 In addition the effective entangling gate depths for each of the circuits are provided in Tab. \ref{tab:gatecostssummary}. From each of these tables, we can see that the qutrit encoding requires in total fewer gates and has shorter circuits that the qubit encoding.

\section{Error models}
\label{sec:Errors}
We look at two well studied \cite{2019arXiv191000471B,2021arXiv210102109G,2020arXiv200108653L,Gokhale_2019,Miller_2018,2013efficienttwirl,nielsen_chuang_2010,2016efficienttwirling,2018efficienttwirling} error models for quantum computing, a Pauli-decoherence channel and a thermal relaxation channel. The explicit representation of these channels are given for both qubit and qutrits but are naturally extendable to higher dimensional qudits.

The Pauli decoherence channel is useful since it is a natural extension of a classical computing error channel. The Pauli decoherence channel is frequently used as the basis for construction of fault tolerant codes \cite{2018efficienttwirling,2013efficienttwirl,2016efficienttwirling,nielsen_chuang_2010}, as well as introductory models to describing noise in quantum simulations \cite{2018arXiv180803623E,PRXQuantum.1.020318}. There is an assumption in the work that follows that single qubit / qutrit gates are approximately $\mathcal{O}(1)$ and noiseless. While this may be functionally true for qudits with some small number of states this may not necessarily true in general for qudits with suppose 20 or more states as the number of rotations scales like  $\mathcal{O}(d^2)$ where $d$ is the number of states. Therefore there is a caveat on this scaling argument presented here; single qudit gates for qudits with a large number of states may not necessarily be noiseless or be functionally instantaneous. 

\subsection{Pauli decoherence channel}
 The Pauli decoherence channel corresponds to some combination of bit-flips ($|0\rangle \leftrightarrow |1\rangle$) which can be modeled by the application of $\sigma_x$ to a qubit and phase-flips ($|1\rangle \leftrightarrow -|1\rangle$), modelled with a $\sigma^z$ happening after a  quantum gate is applied \cite{2018arXiv180803623E,nielsen_chuang_2010,2016efficienttwirling,2018efficienttwirling}. The Pauli-decoherence channel, $\mathcal{E}_{Pauli}^{(2)}$, can be written as
\begin{equation}
	\label{eq:pauli1qubit}
	\mathcal{E}_{Pauli}^{(2)}(\rho) = (1 - p) \rho + p / 4 \sum_{i = 0}^3 \hat{\sigma}_i \rho \hat{\sigma}_i,
\end{equation}
where,$\sigma^i$ are the Pauli group. 
This error channel can be extended to a two qubit one in the following way:
\begin{equation}
\label{eq:pauli2qubit}
\begin{split}
\mathcal{E}_{Pauli-2q}^{(2)}(\rho) = (1 - p_{2,2}) \rho + p_{2,2} / 16 \sum_{i,j}\hat{\sigma}^1_i\hat{\sigma}_j^2 \rho \hat{\sigma}^1_i\hat{\sigma}_j^2.
\end{split}
\end{equation}
The superscripts, $1$ and $2$, indicate the Pauli matrices act on the first and second qubit respectively.

This decoherence channel has a natural extension to qutrits. The single qutrit error channel, $\mathcal{E}^{(3)}_{Pauli}(\rho)$, is
\begin{equation}
\label{eq:pauliqutriterror}
 \mathcal{E}^{(3)}_{Pauli}(\rho)= (1 - p) \rho + \frac{p}{9}\sum_{i=0}^{2}\sum_{j=0}^2 \hat{\mathcal{X}}^i \hat{\textbf{Z}}^j \rho (\hat{\textbf{Z}}^j \hat{\mathcal{X}}^i)^{\dagger}.
\end{equation}
The operators in this noise model are given by the matrices,
\begin{equation}
		\label{eq:qutritpaulis}
		\hat{\mathcal{X}} = \begin{pmatrix} 0 & 1 & 0\\ 0 & 0 & 1\\1 & 0 & 0 \end{pmatrix} \text{ and } \hat{\textbf{Z}} = \begin{pmatrix} 1 & 0 & 0\\ 0 & e^{2i\pi/3} & 0\\ 0 & 0 & e^{4i\pi/3}\end{pmatrix}.
		\end{equation}
This can be modelled as qutrit states being shifted instead of flipped, $\hat{\mathcal{X}}$, and phases being shifted $\hat{\mathbf{Z}}$. 
The 2-qutrit error model is given by a tensor combination of the two. 
\begin{equation}
\label{eq:pauli2qutriterror}
\begin{split}
\mathcal{E}^{(3)}_{Pauli-2q}(\rho) =& (1 - p_{2,3}) \rho + \\&\frac{p_{2,3}}{81} \sum_{i,j,k,l} \hat{\mathcal{X}}_1^i \hat{\mathcal{X}}_2^j \hat{\textbf{Z}}_1^k\hat{\textbf{Z}}^l_2 \rho  (\hat{\mathcal{X}}_1^i \hat{\mathcal{X}}_2^j \hat{\textbf{Z}}_1^k\hat{\textbf{Z}}^l_2)^{\dagger}.
\end{split}
\end{equation}
After every application of the controlled sum gate the density matrix is changed by passing it as the argument to the above expression. Currently values of $p_2$ for qutrits are approximately $0.13$ \cite{morvan2020qutrit}.

\subsection{Amplitude damping channel}
The amplitude damping channel corresponds to idling and circuit time errors and represents excited quantum states decaying to lower energy states. This is typically considered an accurate approximation of some noise on quantum computers \cite{nielsen_chuang_2010,morvan2020qutrit,2013efficienttwirl,2018efficienttwirling,blok:2020may,Gokhale_2019,baker2020efficient} as it is related to the decay time, $T_1$, from the $|1\rangle$ state to the $|0\rangle$ state.  This non-unitary channel for a qubit is given by a the following Kraus map, which describes the evolution of the density matrix \cite{nielsen_chuang_2010}:
\begin{equation}
\label{eq:thermalqubit}
\mathcal{E}_{therm}(\rho) = \hat{K}_0 \rho \hat{K_0} + \hat{K}_1 \rho \hat{K}_1^{\dagger},
\end{equation}
where $K_i$ are Kraus operators of the form,
\begin{equation}
\label{eq:qubitthermops}
\hat{K}_0 = \begin{pmatrix}
1 & 0\\
0 & \sqrt{e^{- t / T_1}}
\end{pmatrix} ~ \text{and } \hat{K}_1 = \begin{pmatrix}
0 & \sqrt{1 - e^{- t / T_1}}\\
0 & 0
\end{pmatrix},
\end{equation}
where $t$ is the implementation time of the quantum gate.
These operators cause the system to decay from the $|1\rangle$ state to the $|0\rangle$. We use this to model the dependence of the system on the simulation time. The generalizations to a qutrit are given by decays from the $|1\rangle \rightarrow |0\rangle$ and $|2\rangle\rightarrow|0\rangle$. The Kraus map for these operators is
\begin{equation}
	\mathcal{E}_{therm}(\rho) = \sum_{i = 0}^{2} \hat{K}_i \rho \hat{K}^{\dagger}_i,
\end{equation}
where 
\begin{equation}
	\label{eq:qutritthermops}
	\begin{split}
	\hat{K}_0 = & \begin{pmatrix} 1 & 0 & 0 \\ 0 & \sqrt{e^{- t / T_1}} & 0\\ 0 & 0 & \sqrt{e^{-2t /  T_1}}\end{pmatrix},\\
	\hat{K}_1 = & \begin{pmatrix} 0 & \sqrt{1 - e^{-t / T_1}} & 0\\ 0 & 0 & 0\\ 0 & 0 & 0 \end{pmatrix}, ~\text{and }\\
	\hat{K}_2 = & \begin{pmatrix} 0 & 0 & \sqrt{1 - e^{-2t/T_1}} \\ 0 & 0 & 0\\ 0 & 0 & 0\end{pmatrix}.
	\end{split}
\end{equation}
While the $T_1$ time for the $|1\rangle\rightarrow|0\rangle$ decay does not have to be one half the $T_1$ time for the $|2\rangle\rightarrow|0\rangle$ decay this is chosen to simplify the parameter space search. 

\begin{figure*}[t]
\centering
\includegraphics[width=\textwidth]{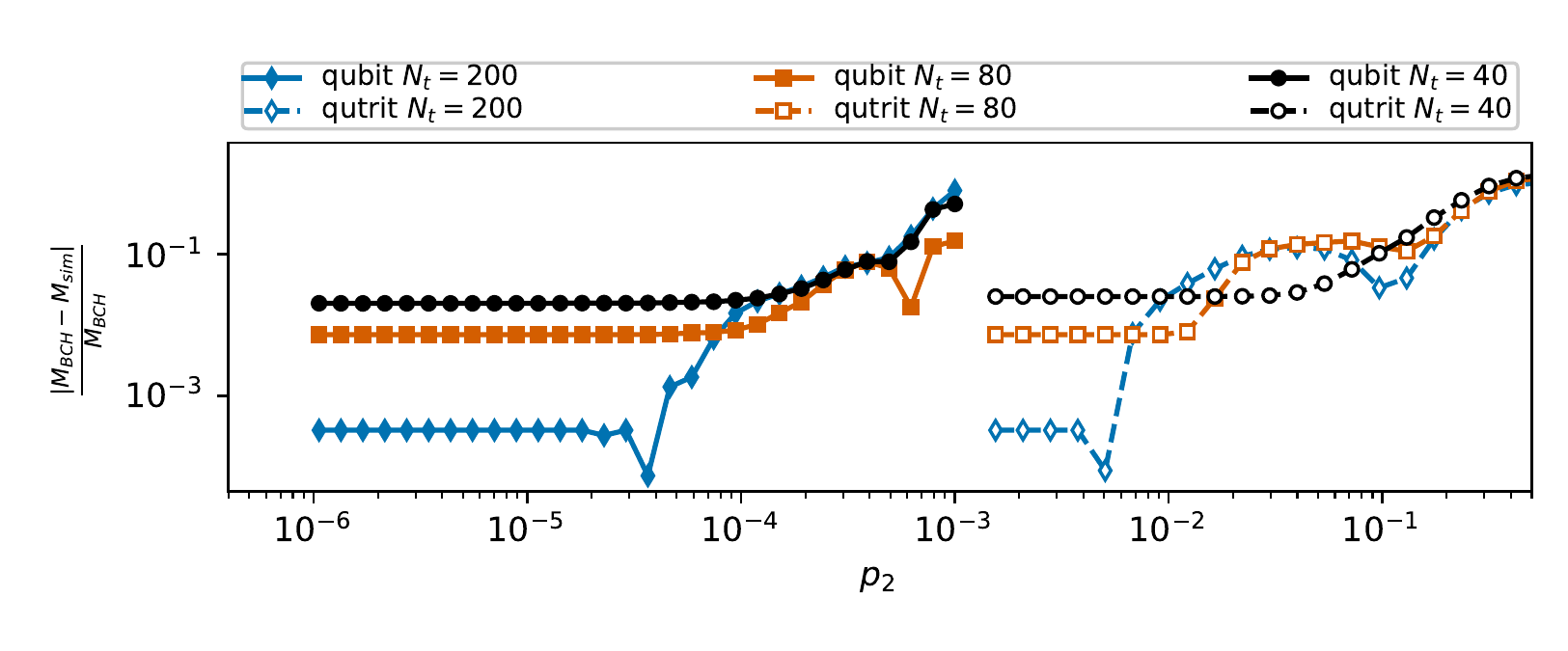}
\vskip-2.5em
\caption{Discrepancy of $m$ from BCH Hamiltonian found via Fourier transform of Eq. (\ref{eq:csQEDcor}) using $N_{ Trotter}N=[40,80,200]$ corresponding to $\delta E_{fft} / m = [20\%, 10\%, 5\%]$ as a function of only the Pauli decoherence error of the qubit (qutrit).}
\label{fig:noamplitudedamping}
\centering
\includegraphics[width=\textwidth]{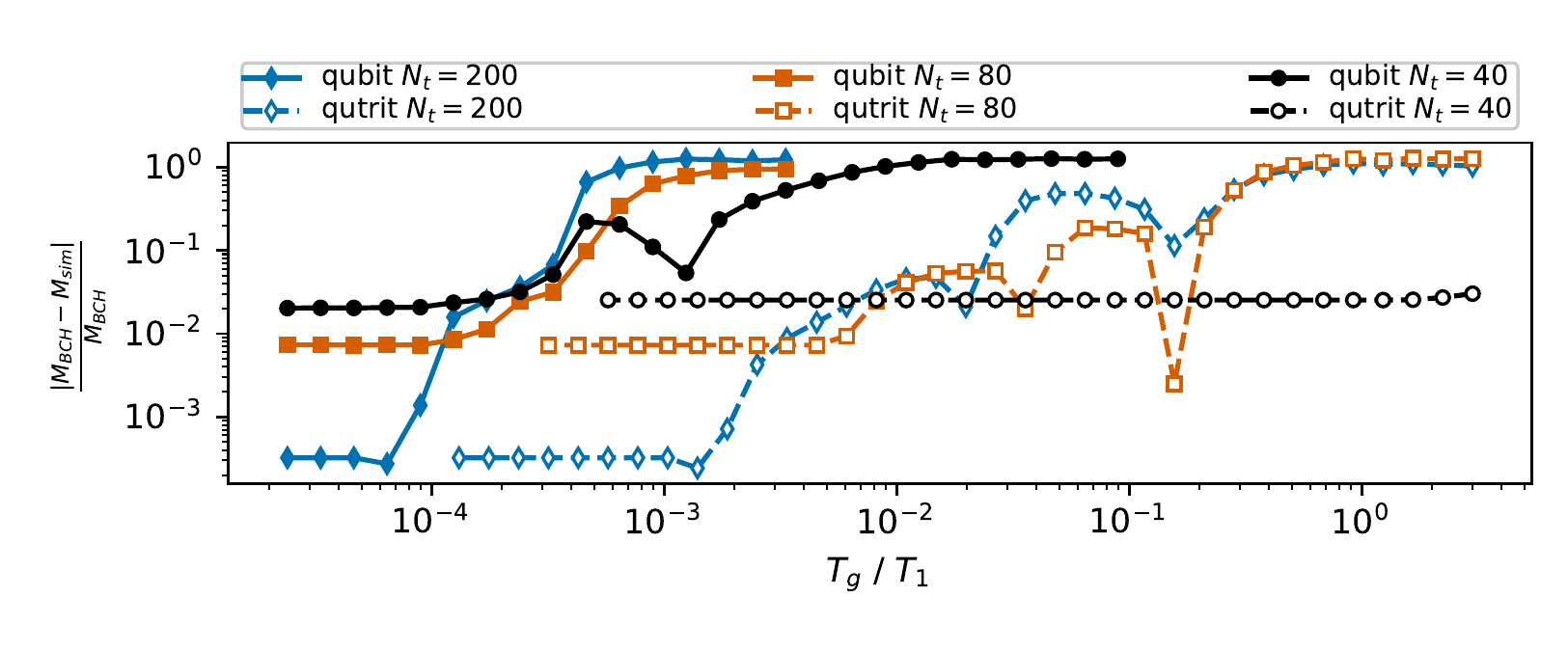}
\vskip-2.5em
\caption{Accuracy of $m$ found via Fourier transform of Eq. (\ref{eq:csQEDcor}) using $N_{ Trotter}N=[40,80,200]$ corresponding to $\delta E_{fft} / m = [20\%, 10\%, 5\%]$ as a function of only the $T_g$ / $T_1$ of the qubit (qutrit) where $T_g$ is the CNOT or CSUM gate time.}
\label{fig:nogateerrors}
\end{figure*}

\section{Results}
\label{sec:Results}

\begin{table}[t]
\begin{tabular}{ccc}
\hline\hline
$\mathcal{A}$ &$p_{2,2} \times 10^{-4}$ & $p_{2,3} \times 10^-2$\\
\hline
20$\%$ & 4  & 8 \\
10$\%$ & 2 & 4 \\
5$\%$ & 1 &1\\
\hline
\end{tabular}
\caption{Estimates for required entangling gate error for qubits and qutrits to achieve  a given accuracy on the mass gap.}
\label{tab:p2soleestimate}

\begin{tabular}{ccc}
\hline\hline
$\mathcal{A}$ & $\frac{T_{CNOT}}{T_1}\times10^{-4}$ & $\frac{T_{CSUM}}{T_1}\times10^{-2}$\\
\hline
20$\%$ & $8 $ & $10$\\
10$\%$ & $8 $ &  $3$\\
5$\%$ & $3$ & $1$\\
\hline
\end{tabular}
\caption{One significant figure estimates of entangling gate-$T_1$ requirements for qubit and qutrit simulations to achieve desired accuracy.}
\label{tab:solet1time}
\end{table}

We discuss the results of using these error models to examine the effects of noise on this quantum simulation. The qubit simulations were performed using QISKit's density matrix simulator \cite{Qiskit}. All the errors are assumed to be contained within the CNOT gate for the Pauli decoherence channel and the amplitude damping channel is applied for all quantum logic operations. The qutrit simulations were implemented in python using numpy for the density matrix representation for the operators and noise channels. The Pauli decoherence channel was applied after the CSUM gate only. While the amplitude damping channel was applied after all $R^x$ and $R^y$ rotations. The simulations used a Trotter step size of $\delta t = 0.235$ and were carried out for $N_{Trotter}$ Trotter steps, with $N_{Trotter} = [40,80,200]$. These choices for $N_{Trotter}$ correspond to accuracies of $20\%,~10\%$, and $5\%$ and are chosen as feasible goals for the near future.   

Before combining the two error models together, we briefly examine general effect both channels have on determinations of the mass independently. The limiting case of Pauli decoherence alone is shown in Fig. \ref{fig:noamplitudedamping}. For both the qubits and qutrit simulations, the noise rate to achieve a precision for $m$ appears to scale as 
\begin{equation*}
    \delta E_{fft} / m \propto p_2.
\end{equation*} where the constant of  proportionality for qubits is $\sim 200$ and the constant for qutrits is $\sim 4$. The best case precision is governed by Eq. (\ref{eq:dE_FFT}), and the required errors $p_{2,2}$ and $p_{2, 3}$ are given in Tab. \ref{tab:p2soleestimate}. In addition we find that the values of $p_{2,3}$ are $\mathcal{O}(10^2)$ larger than for $p_{2,2}$. This indicates that are qutrit machines can be substantially worse and we will still achieve reasonable results with a high fidelity qubit computer.

In the case of solely amplitude damping, we find that the required CNOT gate times need to be less than $10^{-3} T_1$. However the required CSUM gate times need only be less than $10^{-1} T_1$.The required gate times for the different desired precisions. These required gate times for a given precision are listed in Tab. \ref{tab:solet1time} and the accuracy as a function the gate times is shown in Fig. \ref{fig:nogateerrors}. In summary the qutrits $T_1$ can be 100 times greater than the qubit $T_1$ and the same precision can be achieved.

In both of these limiting cases we find that the required gate times and Pauli decoherence error rates are well beyond what is deliverable on current qubit based machines. However, the required gate times and Pauli decoherence rates are much less stringent and are substantially closer to what is deliverable currently. 

\subsection{Combined Noise Model}
\label{sec:detailedstuff}
\begin{figure}[!ht]
\includegraphics[width=0.48\textwidth]{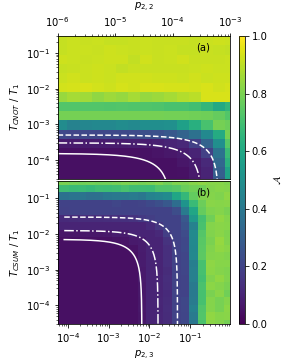}
\caption{Accuracy, $\mathcal{A}$, as a function of entangling gate Pauli decoherence error and 2 qudit gate time for a qubit encoding, (a), and qutrit encoding (b). The dashed white line is the 20 per-cent error threshold. The dot-dashed white line is the 10 per-cent error threshold. The solid white line is the 5 per-cent error threshold.}
\label{fig:correldiff}
\end{figure}

Now that we have examined these asymptotic limits, the combination of these two errors may provide a more robust description of a NISQ machine. Fig. \ref{fig:correldiff} shows the accuracy of the simulations of the qubit and qutrit encodings across several orders of magnitude for $T_1$ time and entangling gate errors. The accuracy is defined as
\begin{equation}
\label{eq:inaccuracy}
\mathcal{A} =\text{max}\Big(\frac{\delta E_m(T_1, p_2)}{E_{m; th.}},\frac{
(E_m(T_1, p_2) - E_{m; th.})}{E_{m; th.}}\Big),
\end{equation}
where $E_{m; th.}$ is the expected mass from Eq. \ref{eq:BCHham}, $E_m(T_1, p_2)$ is the mass derived from the noisy simulation. 
This metric helps combine the accuracy, $\frac{
(E_m(T_1, p_2) - E_{m; th.})}{E_{m; th.}}$, and precision, $\frac{\delta E_m(T_1, p_2)}{E_{m; th.}}$, of the simulation into a single number. 
This inaccuracy in Fig. \ref{fig:correldiff} is shown across the three different evolution times selecting the case among $N_t = 200,~80$, and $40$ that provides the best accuracy. The striking features that we can see from this plot are that across both errors the qutrit simulations achieve similar results with one to two orders of magnitude worse errors than qubit simulations. The lines demarking the estimated required errors in Fig. \ref{fig:correldiff} are provided in Tab. \ref{tab:errorestimates}. 

Its clear that even a four site simulation of this field theory is beyond the capabilities of NISQ era qubit based machines given that for even 20 per-cent accuracy depolarizing errors have to be sub 0.05 percent and $T_1$ times have to be 1000 times longer than the implementation of a CNOT gate. While higher precision (with 5 and 10 percent) are beyond the ability of current qutrit machines \cite{blok:2020may,morvan2020qutrit}, if 20 percent precision is desired it is possible that a NISQ era qutrit machine could extract the mass for a 4 site model as a proof of principle.

\begin{table}[!ht]
\begin{tabular}{ccc}
\hline\hline
$\mathcal{A}$ & qubit error bound & qutrit error bound \\\hline
$20\%$ & 
$\frac{T_{CNOT}}{T_1}\lesssim0.0005-0.8013p_{2,2}$ & $\frac{T_{CSUM}}{T_1}\lesssim0.030-0.618p_{2,3}$\\
$10\%$ & 
$\frac{T_{CNOT}}{T_1}\lesssim0.0003-0.9365p_{2,2}$ & 
$\frac{T_{CSUM}}{T_1}\lesssim0.013-0.774p_{2,3}$\\
$5\%$ & 
$\frac{T_{CNOT}}{T_1}\lesssim0.0002-1.6456p_{2,2}$ & 
$\frac{T_{CSUM}}{T_1}\lesssim0.007-1.091p_{2,3}$\\\hline
\end{tabular}
\caption{Approximate noise rate bounds for qubit and qutrit simulations to achieve a desired accuracy on the mass gap.  $T_g$ is the entangling gate time, $T_1$ is the relaxation time, and $p_2$ is the entangling gate Pauli error}
\label{tab:errorestimates}
\end{table}

\section{Conclusions}
\label{sec:Conclusions}

This work has provided an estimate for the noise requirements for qubit and qutrit machines to simulate a three-state truncation of a csQED on four sites. Its is found that even for four sites qubits will not be feasible for simulating this theory unless extremely high fidelity gates ($> 0.9995$) with extremely fast entangling gates that are approximately 1000 times the $T_1$ time of the qubit. These requirements are an optimistic estimate, if spectator errors and cross talk are additionally included these noise requirements will be even more stringent. For qutrits, the fidelities for the entangling gates can be approximately $0.99$ and scaled gate times can be 1.5 to 2 orders of magnitude shorter than for qubit to achieve approximately the same accuracy as for qubit simulations. 

Nevertheless if these small scale simulations are suggestive, even small-scale problems in $2+1$ and $3+1$ dimensions will likely require some level of error correction. For this reason, it will be important for future work to look at algorithmic costs with similar fault tolerant gate sets to compare the qubit-qutrit-qudit encodings. However if 20 - 100 percent accuracy is acceptable on certain observables, such as transport coefficients \cite{cohen2021quantum}, there maybe a small window where non-error corrected qutrits and qudits may be able to effectively simulate a related $(2+1)$d and $(3+1)$d theory.

\begin{acknowledgements}
I would like to thank Roni Harnik, Henry Lamm, M. Sohaib Alam, Michael Wagman, and Judah Unmuth-Yockey for fruitful conversations and helpful comments.
This work is supported by the DOE QuantISED program through the theory  consortium "Intersections of QIS and Theoretical Particle Physics" at Fermilab and by  the U.S. Department of Energy,  Office of Science,  National Quantum  Information  Science  Research  Centers,   Superconducting  Quantum  Materials  and  Systems  Center (SQMS) under contract number DE-AC02-07CH11359. Fermilab is operated by Fermi Research Alliance, LLC under contract number DE-AC02-07CH11359 with the United States Department of Energy.
\end{acknowledgements}

%

\appendix
\section{Qubit State Preparation Circuit}
\begin{figure*}
\includegraphics[width=6.75in]{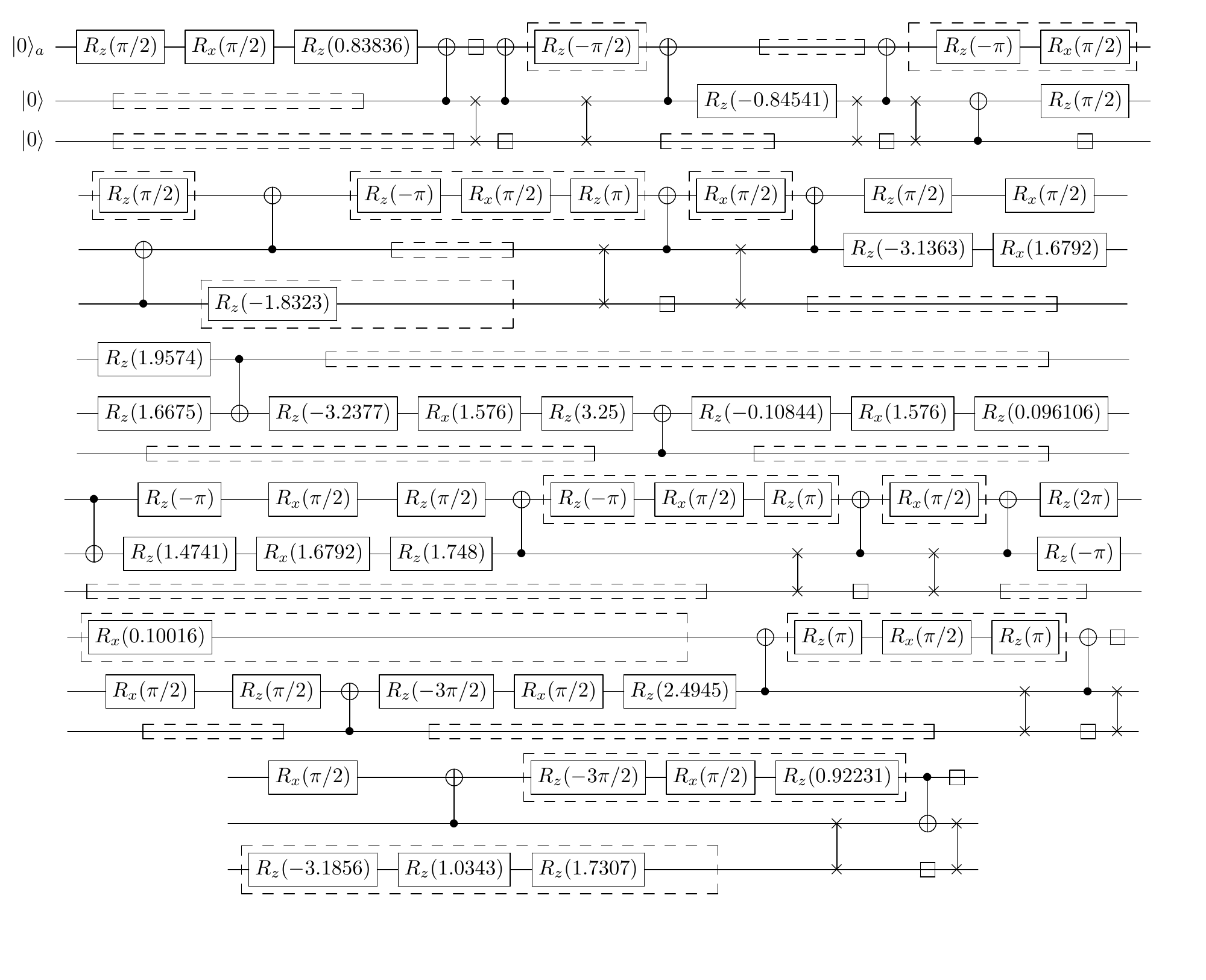}
\caption{Quantum Circuit to implement $\hat{CU}_g^{qbt}$. The dashed regions denote regions where the qubits are idling.}
\label{fig:qubitcug}
\end{figure*}
\end{document}